\documentclass[]{spie}  %>>> use for US letter paper

\usepackage{amsmath,amsfonts,amssymb}
\usepackage{graphicx}
\usepackage[colorlinks=true, allcolors=blue]{hyperref}
\usepackage{booktabs}
\usepackage{authblk}  % 添加这个包来支持 \affil 命令

\title{The Application of Large Language Models in Recommendation Systems}

\author[1]{Peiyang Yu}
\author[1]{Zeqiu Xu}
\author[2]{Jiani Wang}
\author[3]{Xiaochuan Xu}
\affil[1]{Information Networking Institute, Carnegie Mellon University, 5000 Forbes Avenue, Pittsburgh, PA 15213, USA}
\affil[2]{Department of Computer Science, Stanford University, 450 Jane Stanford Way, Stanford, CA 94305, USA}
\affil[3]{Information Networking Institute, Carnegie Mellon University, 5000 Forbes Avenue, Pittsburgh, PA 15213, USA}
\authorinfo{* peiyangy@alumni.cmu.edu}
\begin{document} 
\maketitle

% 摘要部分
\begin{abstract}
The integration of Large Language Models into recommendation frameworks presents key advantages for personalization and adaptability of experiences to the users. Classic recommendation methods, such as collaborative filtering and content-based filtering, are seriously limited in solving cold-start problems, data sparsity, and lack of diversity in the information considered. LLMs, of which GPT-4 is a good example, have emerged as powerful tools that enable recommendation frameworks to access unstructured data sources such as user reviews, social interactions, and text-based content. By analyzing these data sources, LLMs improve the accuracy and relevance of recommendations, thereby overcoming some of the limitations of traditional approaches. This work discusses applications of LLMs in recommendation systems, especially in electronic commerce, social media platforms, streaming services, and educational technologies. This showcases how LLMs enrich recommendation diversity, user engagement, and the system’s adaptability; yet it also looks at the challenges associated with their technical implementation. This can also be presented as a study that shows the potential of LLMs to change user experiences and enable innovation in industries.
\end{abstract}

% 关键词
\keywords{Large Language Models, Recommendation Systems, User Engagement, Candidate Generation.}

% 正文部分
\section{Introduction}
\subsection{Background}
Recommendation lies at the heart of modern digital systems: e-commerce, social media, and streaming. Scientists build these systems to recommend products, content, or people that are the most relevant for users with the aim of improving their user experience. Recommendation systems have used various classical techniques over the last few decades, prominent among which are collaborative filtering and content-based filtering. While useful in certain contexts, both come with major drawbacks. For example, collaborative filtering relies on user interactions and mostly have cold-start issues when new users or new items are added to the system. Similarly, content-based filtering primarily relies on structured data and often misses subtlety in user preferences if data is sparse or non-descriptive. The shortcomings mentioned above create a scope for innovations that may be able to overcome the shortcomings of traditional methods.

This study focuses on applying LLMs in recommendation systems across diverse industries, including B2C e-commerce, streaming platforms, and social media networks. These domains represent significant use cases due to the volume of unstructured data available and their critical need for personalized recommendations.

\subsection{Evolution of Recommendation Systems}
The evolution of recommendation systems has always been guided by continuous innovation, with increasing complexity in user needs and volume. Recommendation systems used to be done mostly through heuristic-based methods, such as collaborative filtering, during the early days, where it identified the pattern in user item interaction matrices for item suggestion. Content-based filtering methods enriched recommendations by bringing metadata and item features into consideration. However, these often failed to resolve issues with regard to the cold start problem, sparsity, and scalability.

Then comes machine learning, a paradigm shift to recommendation systems. Techniques such as matrix factorization, neural networks, and reinforcement learning made possible the powerhouse tools to elicit latent relationships among data. More recently, deep learning introduces the ability of systems to model complex nonlinear interactions between users and items. It has emerged that hybrid systems can combine several methods in order to avoid the weaknesses of a single technique. Recently, large language models have come to play a critical role in this continuously evolving arena by providing unprecedented levels of understanding and generation for human like text, thereby drastically enhancing recommendation quality and personalization.

\subsection{Role of Large Language Models in Recommendation System}
Large language models are a disruptive innovation in artificial intelligence. They achieve state-of-the-art performance in natural language understanding and generation, thanks to large datasets to learn from and deep neural architectures. The capability for processing or understanding huge volumes of textual data opens new avenues for improvement of recommendation systems. While most conventional methods rely on structured user items interactions, LLMs have the capability to derive insight even from unstructured data, such as product descriptions, user reviews, and social media interactions; hence, rich contextual information also enriches the recommendation process. LLMs serve manifold purposes in various recommendation systems: enabling high levels of personalization by elaborate user profiling through textual data, with more precise understanding of user intent from conversational interfaces. LLMs also allow cross-domain recommendations through the connection of insights from diverse datasets with their generalized knowledge. For example, in e-Commerce, LLMs make product suggestions that are in tune with the preferences and interests of users, while in social media and streaming platforms, they allow discovery through the analysis of trends, sentiments, and user engagement patterns. With the integration of LLMs, modern recommendation systems have become more adaptive, context-sensitive, and capable of delivering superior user experiences.

The recent release of large language models, such as GPT-4, have transformed artificial intelligence in the last few years, with little having the reach that now can handle more text than ever. This is because, unlike traditional algorithms, LLMs can automatically process large volumes of unstructured data, including linguistic information such as textual descriptions, user reviews, and conversational exchanges\cite{hadi2024large} Training them on large datasets enables them to catch patterns that allow extremely nuanced recommendations to be generated. They achieve this by analyzing not only user preferences that are appropriately flagged but also more informal latent signals in consideration. For example, LLMs might parse through a user’s past reviews and social media interactions to surface a deeper understanding of their preferences and needs. LLMs have emerged as strong points with substantial strength in knowledge extraction from information and proficiency in user preference understanding. The latter are very much destined to be an important instrument for overcoming the challenge of cold start problems and sparsity in data, thus ushering in new eras for adaptive recommendation engines into several domains.

\subsection{Purpose of the Study}
This survey reviews how large language models have challenged the applications' perspective towards recommendation systems, thereby reviewing its impact. The models open new paths toward recommendation systems by considering the peculiarities introduced by LLMs, among them competence in natural language processing, the capability to capture intricate patterns, and the subtlety of the context\cite{bharathi2024analysis}. The research also removes some mystery from the question of how LLMs can solve some of those perennial problems that have beset traditional systems, namely cold start, data sparsity, and the generally restricted nature of structured data-based analysis. These results are in good agreement with the active experimentation of LLMs within recommendation systems well beyond academia and underline crucially how the friction between ease of onboarding and user experience can inform LLMs' potential to reshape user experiences in e-commerce, streaming, and social networks, among others.
In addition to enhancing the accuracy of the recommendations, the paper discusses higher-order integration of LLMs in light of how these models can enable personalization to a level never seen or experienced. These LLMs are able to generate human-like content and make inferences from unstructured data to deeply understand user preference, behaviors, and intentions\cite{202501.0627}. It is these capabilities that can provide recommendations that are more relevant to the users while at the same time helping build trust and engagement. This paper further discusses the key challenges during LLM integration: computational requirements and domain specific fine tuning. It discusses challenges, while also providing a pointer toward future research in this domain; hence, the work may be of interest for both researchers and practitioners by realizing the full potential of LLMs for the development of next-generation recommendation systems.

\section{Fundamental Concepts of Recommendation Systems and Large Language Models}

\subsection{Recommendation System Overview}
Recommendation systems are complex algorithms that suggest personalized items to a user by predicting the most relevant items of interest to the user. They analyze user behavior, preferences, and interaction history and then provide personalized recommendations. Recommendation systems have evolved over time from a 'nice to have' feature to a core component of digital platforms, ranging from e-Commerce and social networking to streaming services and online learning. Some of the major techniques applied in recommendation systems include:

\begin{itemize}
    \item Collaborative Filtering (CF): This method makes predictions by observing the patterns of similarity between different users or items using user item interaction data. Using these patterns, the system will predict the likelihood of a user liking an item based on the preferences of similar users, known as user-based CF\cite{widayanti2023improving}, or by highlighting items that are commonly consumed together, known as item-based CF.3 Yet, collaborative filtering usually suffers from data sparsity and cold start problems when interaction data from either users or items is limited.
    
    \item Content-Based Recommendation: Unlike CF, this approach focuses on the intrinsic properties of items. It utilizes item attributes, such as genre, keywords, or features, to recommend similar items to those that a user has liked or with which they have previously interacted. While effective in some applications, content-based methods can lack diversity and often fail to uncover novel recommendations beyond the user’s existing preferences.
    
    \item Hybrid Methods: Hybrid methods overcome some of the limitations of CF and content-based approaches by combining their merits. Hybrid models that integrate collaborative and content-based filtering techniques enhance the accuracy of recommendations, promote diversity, and alleviate some issues such as cold starts and sparsity.
\end{itemize}

Mathematically, let \( R_{u,i} \) represent the recommendation score for a user \( u \) and an item \( i \). This can be expressed as:
\[
R_{u,i} = f(u, i) + \epsilon,
\]
where \( f(u, i) \) represents the prediction model derived from past interactions, and \( \epsilon \) is an error term that accounts for uncertainties or deviations in prediction.

\subsection{Overview of Large Language Models}
Large language models, such as GPT-4, represent a paradigm change in natural language processing and AI-driven applications. These models are based on transformer architectures and pre-trained on large datasets comprising unstructured text from a wide variety of domains. The intrinsic strength of these models is to generate, understand, and manipulate natural language; therefore, they can be very versatile for translation, summarization, and recommendation tasks. In the development of LLMs, there mainly exist two major phases:

\begin{itemize}
    \item Pre-Training: During this stage, the model is pre-trained on large, unlabeled datasets to grasp the language pattern, grammatical structure, and contextual relationship. Pre-training allows it to abstractly have a basic view of the language by which it generalizes over most of the tasks easily\cite{awais2023foundational}.
    
    \item Fine-tuning: After pre-training, the model is fine-tuned on domain-specific data with labels pertaining to some application. It refines the knowledge of the model and aligns its output for specialized tasks, say recommendation generation or sentiment analysis.
\end{itemize}

LLMs leverage attention mechanisms\cite{yi2024optimization} that allow them to capture intricate contextual relationships between words, phrases, and sentences. That is what makes them great for recommendation systems that involve deep insight into user preferences, mostly conveyed by natural language. For example, LLMs can analyze user reviews, extract sentiment, and find nuanced preferences leading to highly personalized, contextually relevant recommendations. This will involve embedding LLMs into recommendation systems and enabling them to go beyond structured data to unlock the power of unstructured inputs.

\section{Framework for Applying Large Language Models in Recommendation Systems}

Large language models should be integrated into recommendation systems by designing a robust framework that manages data preprocessing, candidate generation, personalized ranking, and multimodal fusion. This section describes an overall framework composed of modular components that define processes and methodologies that enable LLMs to provide highly personalized and relevant recommendations.

\subsection{Data Preprocessing and Input Formats}
Data preprocessing is the basis to use LLMs in recommendation systems, which transforms diversified forms of user, item, and interaction data into formats suitable for LLMs to process effectively.

Text data will include product descriptions, user reviews, and item metadata; this text-based information needs to be tokenized and then embedded. Each of these embeddings bears semantic information encapsulated in a vector space preserving contextual relationships. For a given product description $di$, the tokenized output $T(di)$ is represented as:
$$
E(di) = \text{Embedding}(T(di))
$$
where $E(di)$ is the embedding vector of the product description.

User behavior data, such as clickstreams, purchase histories, and preferences, is summarized into a feature vector $X_u$:
$$
X_u = [E(di_1), E(di_2), \ldots, E(di_n)]
$$
Here, $di_1, di_2, \ldots, di_n$ are the descriptions of items the user has interacted with.

Other features extracted from raw data to augment the input of LLMs include sentiment scores, key phrases, or temporal patterns. These can be either appended to embeddings or used separately:
$$
X_u = [X_u, \text{Sentiment}(u), \text{Temporal}(u)]
$$
Different forms of unstructured data may be used to better the recommendation systems: reviews by users on Amazon and Yelp, social media interaction through posting on Twitter and comments, history from Spotify and Netflix. User reviews capture the sentiment and granular product feedback, while social media captures trends and user preferences in near-real time. Streaming histories provide context for user consumption patterns and preferences. These sources of data were collected using API integrations and web scrapers, ensuring that any platform specific policy was respected in the process. Furthermore, processing of such diverse unstructured data allows LLMs to generate richer recommendations with personalization, thus alleviating issues related to cold start problems or data sparsity.

\subsection{LLM Training Process}
The training of LLMs is a two-phase process: pre-training and fine-tuning. During pre-training, the models are exposed to large, diverse datasets such as Common Crawl and Wikipedia for wide knowledge about pattern, context, and semantic relations\cite{ma2024fake}. Fine-tuning then occurred with domain-specific datasets, including user reviews from sites like Yelp, product descriptions from e-commerce databases, and interaction logs from social media and streaming services. In such a way, the fine-tuning stage aligned the LLMs with the specific tasks of sentiment analysis, preference extraction, and contextual recommendation generation. Besides, data augmentation techniques were performed to alleviate cold start and sparsity issues, such as synthesizing user preferences using LLM-generated synthetic reviews. The performance of these LLMs was compared using ranking metrics such as NDCG and MRR, showing a significant improvement over traditional methods in personalization and relevance.

\subsection{Candidate Generation and Recommendation Strategies}
Candidate generation is a very crucial step in narrowing the huge pool of items to a manageable set of relevant options for personalized ranking. LLMs contribute by leveraging both structured and unstructured data in generating candidates.

LLMs embed both users and items in one common vector space and calculate similarity scores to attain semantic matching:
$$
S(u,i) = \text{cosine}(E(xu), E(di))
$$
Items with the highest similarity scores are selected as candidates.

LLMs use contextual understanding of unstructured data, such as reviews, to predict user preference and refine candidate pools through dialogue.
$$
X_u^{\text{updated}} = \text{LLM}(X_u, \text{User\_Response})
$$
This iterative process will help ensure candidates are aligned with the user's evolved preferences.

Diverse candidate generation ensures varied recommendations by selecting candidates from multiple clusters.
$$
C(u) = \bigcup_{k=1}^K \text{Top}(S_k(u,i), n_k)
$$
where K is the number of clusters and $n_k$ is the number of top items selected from each cluster.

\subsection{Personalized Ranking}
If there is a generation of the candidate list, the personalized ranking will score them with respect to user preference and contextual relevance. Such signals include historic interactions, real-time behaviors, and situational context to assess if the ranking is reflecting the individual's need. Such continuous adaptation with respect to implicit and explicit user feedback will refine the system recommendations toward accuracy and engagingness. Thus, the approach transforms the candidate list into a prioritized selection that maximizes user satisfaction.

The ranking model gives scores to candidates with the help of features based on user interactions and item characteristics.
$$
y^u = \text{Rank}(\text{LLM}(xu))
$$
where $y^u$ represents the ranked list of items for user $u$.

Dynamic systems incorporate real-time user interactions, such as clicks and queries, directly into the ranking process:
$$
xu_{\text{final}} = [xu, \text{RealTime}(u)]
$$

$$
y^u = \text{Rank}(\text{LLM}(xu_{\text{final}}))
$$

\subsection{Multimodal Fusion}
Multimodal fusion\cite{yu2024applications} is a technique that merges data streams from multiple sources, text, images, and audio, into one stream to enhance the quality of recommendations. It enables the detection of various aspects of user preferences and item characteristics.

Multimodal features are combined into one unified representation that improves the ranking, putting together data from various sources such as text, images, user interactions, and contextual signals. This fusion enables the system to capitalize on diverse types of information, capturing complex relationships and patterns that improve the accuracy and relevance of the rankings. This unified feature allows the ranking model to make more informed and holistic decisions in an effort to provide highly personalized and effective recommendations.
$$
X_{u_{\text{multi}}} = [X_u, \text{Image}(i), \text{Audio}(i)]
$$
where $\text{Image}(i)$ and $\text{Audio}(i)$ are embeddings of visual and auditory data, respectively.

\subsection{Evaluation Metrics}
The performance of the trained LLM models in the recommendation system was evaluated using several standard metrics commonly used in the recommendation systems to measure accuracy, ranking quality, and user satisfaction. Key metrics included
\begin{itemize}
    \item Normalized Discounted Cumulative Gain (NDCG): it evaluates the ranking quality by considering the positions of relevant items in the recommendation list;
\item Mean Reciprocal Rank (MRR): it assesses the rank position of the first relevant item for a user;
\item Precision@K: it measures the proportion of relevant items in the top K recommendations
\item Click Through Rate (CTR): it reflects user engagement by calculating the percentage of recommended items clicked. 
\end{itemize}

These metrics collectively provided a comprehensive view of the LLM's performance across different aspects of recommendation quality, allowing comparisons with baseline models and traditional techniques.

\section{Applications of Large Language Models in Recommendation Systems}
Large Language Models are creating significant novelties for Recommendation Systems, making such systems personalized and relevantly engaged in several different domains. On the other hand, their much better processing results in enhancing recommendations’ accuracy to fit naturally in a concrete user setting.

Specific and more detailed descriptions regarding applied LLM scenarios are included hereinafter referencing key areas as follows.

\subsection{E-commerce Recommendations}
In this fast-growing e-Commerce world, effective recommendations and increased customer satisfaction are indeed the drivers of sales. Using LLM, large volumes of unstructured data such as product reviews, descriptions, and customers’ feedback can be analyzed to gain insights into user preference. Whereas the traditional model will return a very neat and structured interaction data table of reviews and professor sentiment toward the underlying features of the product quality, pricing, design, etc., LLMs mine the review text for sentiment, tone, and hotwords defining user sentiments. Assume that an LLM is going to understand from the details in the review or even search queries made by the user that the user is concerned about environmentally friendly technology. From this, the system learns the customer's preferences, since often the customer may not even have searched for them specifically. Moreover, LLMs will also allow real time adjustability: they make better suggestions as the actions become more recent, including items thrown into a cart and searches that have gone astray. Active potential here adds to greater engagement, as recommendations remain relevant throughout each step of the customer journey. Also, LLMs can allow e-Commerce to detect trends in big data and further make necessary changes in marketing campaigns or inventory management strategies for businesses.

Besides, the ability of LLMs to analyze feedback from various sources, such as social media or customer service interactions, offers deeper insights into consumer sentiments. These may be emerging trends related to sustainable product preferences or a change in priorities for customers. E-commerce platforms use this insight into the change to continuously adapt and update their offerings and recommendations in real time. Continuous learning ensures recommendations are not only personalized but also tuned to the changing market trends for higher customer satisfaction and better conversion rates. LLMs also allow companies to discover opportunities for product improvement or extension by finding out what customers need and do not get, hence the opportunity to introduce new products or services that will be well accepted by the target audience\cite{nasseri2023applications}.

Integration of LLM with other technologies, some examples include recommendation algorithms and predictive analytics, provides broader avenues in targeting. Information such as product reviews and browsing activities, captured by LLMs, could form the input to enable marketers to recommend products based on user preference and use his algorithmic knowledge to determine user purchasing behavior over a coming number of weeks. This level of personalization, much more than mere recommendations, makes the shopping experience alive and active. Adopting ever-improved LLMs through a user’s journey lets businesses build deeper relationships with their customers than any competitor, and that means for customers, they want to spend more time with them, resulting in better retention rates that drive more sales over time.

\subsection{Social Media Recommendations}
Social media platforms derive much value from user interaction; therefore, personalized content and connection recommendations become vital. The value of LLMs in this context is to make predictions about the interest and preference of users by analyzing unstructured UGC, which can be in the form of posts, comments, likes, and shares\cite{piriyakul2023automated}. Unlike most of the earlier systems that considered only a few interaction metrics, LLMs delve deeper into the semantic context of UGC and find much more fine-grained themes and interests.

For example, it will be a model that shall infer from the comments a user makes and shows interest in, say in sustainable living or emerging technologies. It will therefore recommend reading materials, videos, or groups for the user that best suit such interests. LLMs can also dig deeper into the patterns of sentiment and engagement in interactions between users to ensure that content recommendations are relevant and constructively impactful. Beyond content recommendations, LLMs can facilitate meaningful social connections in suggesting friends, real news\cite{yu2021text}, or communities based on shared values or mutual interests\cite{ratican2024advancing}. This creates a feeling of belonging and further motivates users to spend more time on the platform. Moreover, their capability for real-time trend adaptation allows platforms to push timely, contextually relevant content, amplifying user engagement with viral events or breaking news.

More importantly, the real-time adaptability of LLMs will keep recommendations in tune with the dynamics of changing user preferences. If a user becomes more engaged in posts on trendy social issues or popular TV series, for instance, the system should immediately adapt to change recommendations towards similar content that keeps the user interested in novelty. This dynamic capability enhances user satisfaction and contributes to higher retention rates, as users are likely to remain attached to a platform offering them relevant and interesting content matching their ever-changing preferences. The constantly updated recommendations make for a more engaging and personalized user experience, making the platforms feel more intuitive and responsive to the needs of the individual.

\subsection{Streaming Recommendations}
Services like Netflix, Spotify, and YouTube are all deeply reliant on title recommendations to retain viewership and time spent watching or listening. LLMs bring a new level of sophistication to these systems by analyzing unstructured data, such as user reviews, viewing or listening histories, genres, plot descriptions, or song lyrics\cite{taief2024application}.

For instance, take a user whose primary genre of watched movies is dystopian sci-fi. In that case, an LLM will be better positioned to suggest some of those very esoteric titles that share similar themes by comparing summaries of plots for similar narrative elements. This also extends to audio platforms where LLMs will study the lyrical content, mood, or song user playlists to create customized music recommendations or theme-based playlists. In addition, LLMs can track changing user preferences, maybe seasonal, say holiday music or taste changes, and provide recommendations for those changing preferences. The LLMs make sure that recommendations through explicit user actions like ratings or likes with implicit signals like time spent on a genre are accurate but also in tune with the current mood and context of the user. It allows the satisfaction of user preferences to a higher level and provides support for their loyalty in the long term.

In addition, LLM can foresee certain user behaviors based on the development trends of user preferences. As an example, some users prefer to hear only songs sung by one kind of artist and gradually start to switch to other categories over time. Then it can pick up on this and continue to provide other artists or songs in the same mold as the user's musical preferences expand\cite{gao2024aligning}. Continuously learning from user interactions and contextual factors, the ever more personalized ways to keep users engaged and interested in new discoveries are through LLMs. This dynamic capability for personalization of recommendations in real time seals that edge for platforms in the ever-changing digital landscape of entertainment today. Since LLMs change with ever-changing user preferences, their content remains fresh and relevant to retain users in the best way possible. These will make them very useful tools for platforms with a view toward increasing user engagement and satisfaction.

\section{Key Technical Challenges}

Although LLMs provide substantial benefits to recommendation systems, they also present several technical and operational challenges. These challenges must be addressed to ensure their seamless integration and maximize their impact.

\subsection{Real-Time Efficiency}

Computational intensity poses a significant hurdle for real-time LLM-based recommendation systems. The resource-intensive nature of LLM predictions can lead to increased latency, which is particularly detrimental in applications like e-commerce where prompt suggestions need to be generated instantly. To mitigate this, several promising strategies can be employed.

\begin{itemize}
    \item Model Optimization Techniques: Techniques like model pruning, which removes redundant connections, quantization, which reduces the precision of model weights, and knowledge distillation, which transfers knowledge from a large model to a smaller one, can significantly reduce model size and computational complexity while maintaining performance.
    
    \item Hybrid Architectures: Combining lightweight models for real-time inference with more powerful LLMs for in-depth offline analysis offers a compelling approach. Lightweight models can handle time-sensitive tasks like candidate generation, while LLMs provide refined suggestions based on deeper analysis.
    
    \item Hardware acceleration: Leveraging specialized hardware like GPUs and TPUs can significantly accelerate LLM inference. These accelerators are designed to handle the parallel computations required by LLMs, leading to substantial speedups. Furthermore, research on energy-efficient hardware is crucial for sustainable LLM deployment in resource-constrained environments.

    \item Algorithmic Innovations: Exploring novel algorithms that can efficiently utilize specialized hardware and optimize energy consumption is essential. This includes developing algorithms that can process data more efficiently and reduce latency while maintaining recommendation quality.
\end{itemize}

By effectively addressing these challenges through a combination of model optimization, hybrid architectures, hardware acceleration, and algorithmic innovations, we can ensure that LLM-based recommendation systems are both efficient and responsive, providing a seamless user experience.

\subsection{Cold Start and Data Sparsity}
The two basic problems that any RS might have are probably cold start problems and sparsity data. Cold-start problems arise when a model requires a sufficient amount of interaction data from both new users or new items in order to make an accurate recommendation; this is essentially incomplete user-item interaction matrices, whereby it is hard to find patterns in the data. LLMs offer several innovative solutions to overcome these limitations.

\begin{itemize}
\item Leveraging External Data: LLMs can effectively use external data sources such as user reviews, social media posts, and product descriptions to infer user preferences and item characteristics. By analyzing these rich textual data, LLMs can generate more accurate recommendations, even with limited interaction data.

\item Synthetic Data Generation: LLMs can generate synthetic data, such as simulated user reviews or preferences, to augment sparse datasets. These synthetic data can be used to train recommendation models and improve their performance in scenarios with limited historical interactions.

\item Real-time Feedback and Adaptation: LLMs can continuously learn and adapt based on real-time user feedback and interactions. By analyzing user behavior patterns, such as clicks, purchases, and ratings, LLMs can refine recommendations dynamically and provide more personalized experiences, even for new users.

\item Collaborative Filtering with LLMs: Integrating LLMs with traditional collaborative filtering techniques can improve the precision of the recommendations. LLMs can provide richer user and item representations by analyzing textual data, which can then be incorporated into collaborative filtering algorithms to improve the quality of recommendations.

\end{itemize}

By effectively leveraging external data, generating synthetic data, and incorporating real-time feedback, LLMs can mitigate the challenges of data sparsity and cold start problems, leading to more accurate and personalized recommendations for all users.

\subsection{LLM Integration}
Integration of LLM into existing recommendation systems is not without challenges. Technically, the integration of LLMs may require major architectural changes to existing systems that would need adaptation for the increased computational requirements of LLMs, probably hardware upgrades, or the adoption of more efficient model-serving strategies. Other operational challenges include ensuring data quality and addressing privacy concerns. Robust data governance practices should be in place while integrating various sources of data into LLMs for the accuracy, fairness, and relevance of the data. Additionally, due consideration should also be given to ensuring user privacy, since LLMs are likely to process sensitive user data. To mitigate these risks and earn user confidence, stronger privacy-preserving mechanisms such as differential privacy and federated learning must be employed.

\section{Conclusion}
\subsection{Key Findings}
Large Language Models represent a big change in recommendation systems, providing great personalization, dynamic interactions, and contextually aware suggestions. These models work best by extracting insights from unstructured data on user preferences that were not easy to address by traditional recommendation systems due to cold start problems and data sparsity. This versatility has also been demonstrated across a number of domains, including e-commerce, social networks, and streaming platforms, increasing user engagement and satisfaction with a service.

\subsection{Research Gaps}
Despite the potentially transformative power of LLMs, there are a series of challenges related to real-time efficiency and overgeneralization. Those have prevented seamless integration into recommendation systems so far. The current workarounds, model fine tuning and incorporating real-time feedback, do only partial justice. Scalability in real-life applications also needs better computational resources.

\subsection{Future Directions}
Advancements of multimodal learning create a whole new dimension for future recommendation systems\cite{wang2024llm}. Techniques such as TriChronoNet, hyper-relational interaction modeling in multimodal trajectory prediction, and Modality Fusion Vision Transformer offer a whole new way of integrating diverse data modalities, ranging from text, images, and audio to user behavior. For example, adding visual aspects to a fashion recommendation system along with queries from the user will yield more accurate, yet engaging suggestions. As the field evolves, the integration of LLM with multimodal learning will probably redefine personalized experiences, giving way to much more sophisticated recommendation systems.

% References
\bibliography{report} % bibliography data in report.bib
\bibliographystyle{spiebib} % makes bibtex use spiebib.bst

\end{document}